\documentclass{article}
\usepackage[dvips]{graphicx}
\usepackage{amsmath}
\usepackage{epsfig}
\newcommand{\bfr}{\begin{flushright}}
\newcommand{\efr}{\end{flushright}}
 
\begin{document}
\title{A New $2+1$ Dimensional Gravity Solution Coupled to Non Linear
Electrodynamics with a Cosmological Constant
%\thanks{Presented at ...}%
}
\author{Masashi Kuniyasu\footnote{s007vc@yamaguchi-u.ac.jp}
\\
%\address
{\small
Graduate School of Science and Technology, Yamaguchi University,}\\
{\small Yamaguchi-shi, Yamaguchi 753--8512, Japan}
}
\date{\today
}
\maketitle
\begin{abstract}
A solution for the Einstein gravity coupled with non linear
electrodynamics is introduced in $2+1$ dimensions. Especially, in the
case with a non-vanishing cosmological constant, we obtain a novel
black hole solution. To find fundamental characters of the solution,
radial null rays in the space-time described by the solution are
investigated.
\\
~
\\
%PACS number(s): ????
\end{abstract}
%\PACS{}

%%%%%%%%%%%%%%%%%%%%%%%%%%%%%%%%%%%%%%%%%%%%%%%%%%%%%%%%%%%%%%%%%%
\section{Introduction}
%%%%%%%%%%%%%%%%%%%%%%%%%%%%%%%%%%%%%%%%%%%%%%%%%%%%%%%%%%%%%%%%%%
The most standard theory of gravitation is Einstein's general theory
of relativity in $3+1$ space-time.
However, since general theory of relativity is classical theory,
this theory needs to be quantized.
This problem is one of the most biggest issue in modern physics.
There are many consideration to quantize the theory of gravitation
in the past half century.
Especially, more simpler theory like a $2+1$ Einstein gravity has
been considered since around 1980's.
They are easier than higher dimensional ones, so it
is hoped that it will be quantized.

The advantages of the lower dimensional gravity are to be able to 
obtain exact solutions easier than the higher dimensional one.
In addition, $2+1$ gravity has some characteristic structure.
First, the Weyl tensor always vanishes in $2+1$ dimensions.
Second, there is no propagating mode
in $2+1$ dimensional Einstein gravity.
Besides, geometry itself differs from that in the higher dimensional
case; for instance, the asymptotic behavior of the
black hole space-time is peculiar in $2+1$ dimensions. The review article
\cite{SC} summarized study on $2+1$ dimensional gravity.

Gravity theory in $2+1$ dimensions came to attract much attention,
especially after a black hole solution has been found.
This black hole is called BTZ black hole\cite{BTZ}, and it is in the
space-time that presence of a negative cosmological constant $\Lambda<0$.
In the limit of zero mass and zero angular momentum, there appears a naked
singular point. The geometric structure and thermodynamical property of
the BTZ space-time is also well investigated.

We note that the existence of a black hole solution in $2+1$ dimensional
world attract much attention to theory of $2+1$ gravity. That is why to
find a new solution in $2+1$ dimensional Einstein gravity is important.
In fact, recently new solution is considered in
the case with other classical matter fields. As the classical field,
a real scalar field \cite{SS}, a field of nonlinear
electrodynamics (NED) \cite{MH}(for example, the Born-Infeld type 
electrodynamics \cite{BI} is one of the NED, which attract much 
attention in string theory \cite{Sav}), and the other fields are adopted
and the solution with circular symmetry is investigated.
The common feature of these substances is a non
vanishing radial component of the stress tensor $T_r^r$. 
The authors of Ref.~\cite{MH} claim that the solution with
non-vanishing cosmological constant and  NED describes a flat space-time
because of the stress tensor from NED and the cosmological constant cancel out.
However, component of the stress tensor from NED is $T^r_r$,
on the other hand, the cosmological constant generates the stress tensor
not only $T^r_r$ but also $T^t_t$ and $T^{\theta}_{\theta}$. 
That is why we doubt the result of Ref.~\cite{SS} with the cosmological constant.

In the present paper, we derive a new solution for $2+1$ gravity with
non-vanishing cosmological constant and NED.
In this paper, we use the natural system of unit. In addition, we consider
the space-time metric with the signature $(-++)$.

%%%%%%%%%%%%%%%%%%%%%%%%%%%%%%%%%%%%%%%%%%%%%%%%%%%%%%%%%%%%%%%%%%
\section{NED Lagrangian and Einstein field equation in $2+1$
dimensional gravity
\label{sec2}}
%%%%%%%%%%%%%%%%%%%%%%%%%%%%%%%%%%%%%%%%%%%%%%%%%%%%%%%%%%%%%%%%%%
The following action which governs the Einstein gravity and NED
is considered:
\begin{equation}
I=\frac{1}{2}\int
d^3x\sqrt{-g}\left(R+\frac{2}{\ell^2}+\alpha\sqrt{|{\cal F}|}\right)\,,
\label{eq1}
\end{equation}
where ${\cal F}=F_{\mu\nu}F^{\mu\nu}$, $F_{\mu\nu}=\partial_{\mu}A_{\nu}-
\partial_{\nu}A_{\mu}$ is the Maxwell invariant, $\alpha$ is a coupling
constant and $\ell$ is related to the cosmological constant $\Lambda$
with $ \ell = \sqrt{-1/\Lambda}$.
Let us take a circular symmetric metric ansatz:
\begin{equation}
ds^2=-A(r)dt^2+B(r)dr^2+R(r)d\theta^2\,,
\label{eq2}
\end{equation}
where $A(r)$, $B(r)$, $R(r)$ are arbitrary functions depending on $r$.
Next, we assume that the field strength tensor $F_{\mu\nu}$ is
calculated from the vector potential ${\bf A}$
\begin{equation}
{\bf A}=E_0 (-b\theta, 0, at)\,,
\end{equation}
where $a$, $b$ are constants. In Ref.~\cite{MH}, they are taken as
$a+b=1$. We thus take the same condition in the present paper.
Under this condition, non-vanishing component of $F_{\mu\nu}$ is
\begin{equation}
F_{t\theta}=\partial_{t}A_{\theta}-
\partial_{\theta}A_{t}=E_0
\label{eq4}
\end{equation}
and
\begin{equation}
F^{t\theta}=g^{tt}g^{\theta\theta}F_{t\theta}=-\frac{E_0}{A(r)R(r)}\,.
\label{eq5}
\end{equation}
From Eqs.~(\ref{eq4}) and (\ref{eq5}), ${\cal F}$ becomes
\begin{equation}
{\cal
F}=2F_{t\theta}F^{t\theta}=-\frac{2E_0^2}{A(r)R(r)}\,.
\label{eq6}
\end{equation}

Einstein field equation derived from the action (\ref{eq1}) is
written as
\begin{equation}
G_\nu^\mu+\frac{1}{\ell ^2}\delta_\nu^\mu=T_\nu^\mu\,,
\end{equation}
where $G_\nu^\mu$ is the Einstein tensor defined as
$G_\nu^\mu=R_\nu^\mu-\frac{1}{2}R\delta_\nu^\mu$ and $T_\nu^\mu$ denotes the
stress tensor
\begin{equation}
T_\nu^\mu=\frac{\alpha}{2}\sqrt{|{\cal F}|}\left(\delta_\nu^\mu
-\frac{2F_{\nu\lambda}F^{\mu\lambda}}{{\cal F}}\right)\,.
\label{eq8}
\end{equation}
If $\mu\ne\nu$, Eq.~(\ref{eq8}) indicates that $T^{\mu}_{\nu}$ vanishes. 
Let us write the diagonal components of the stress tensor explicitly.
They are found to be
\begin{equation}
T_t^t=T_\theta^\theta=\frac{\alpha}{2}\sqrt{|{\cal
F}|}\left(1-\frac{2F_{t\theta}F^{t\theta}}{2F_{t\theta}F^{t\theta}}
\right)=0\,,
\end{equation}
\begin{equation}
T_r^r=\frac{\alpha}{2}\sqrt{\frac{2E_0^2}{A(r)R(r)}}\,.
\end{equation}
Therefore, only non-zero component of the stress tensor is $T_r^r$.
It is the same result as the case with a real scalar field, which
depends only on the radial coordinate \cite{SS}.

The non-zero components of the Einstein tensor obtained from the metric
(\ref{eq2}) are 
\begin{equation}
G_t^t=\frac{2R''BR-R'{}^2B-R'B'R}{4B^2R^2}\,,
\end{equation}
\begin{equation}
G_r^r=\frac{A'R'}{4ARB}\,,
\end{equation}
\begin{equation}
G_\theta^\theta=\frac{2A''BA-A'{}^2B-A'B'A}{4B^2A^2}\,,
\end{equation}
where ${}'$ means the derivative with respect to the radial coordinate $r$.

Finally, we get the field equation as follows:
\begin{equation}
\frac{2R''BR-R'{}^2B-R'B'R}{4B^2R^2}-\frac{1}{\ell^2}=0\,,
\label{eq14}
\end{equation}
\begin{equation}
\frac{A'R'}{4ARB}-\frac{1}{\ell^2}=\frac{\alpha}{2}\sqrt{\frac{2E_0^2}{A(r)R(r)}}\,,
\label{eq15}
\end{equation}
\begin{equation}
\frac{2A''BA-A'{}^2B-A'B'A}{4B^2A^2}-\frac{1}{\ell^2}=0\,.
\label{eq16}
\end{equation}
As seen from Eqs.~(\ref{eq14}) and (\ref{eq16}), 
we can set $A=KR$, where $K$ is a constant. Since $K$
is taken to be unity in Ref.~\cite{MH}, we also take $K=1$ in the
following discussion. Then, Eq.~(\ref{eq15}) becomes
\begin{equation}
\frac{A'{}^2}{AB}-\frac{4}{\ell^2}
A=\frac{4\alpha E_0}{\sqrt{2}}\,.
\label{eq17}
\end{equation}
In Ref.~\cite{MH}, the corresponding term to the second term in
Eq.~(\ref{eq17}) seems to be misplaced. Therefore they obtained the wrong
solution for $\Lambda \neq 0$.

On the other hand, the condition $A=R$ makes Eqs.~(\ref{eq14}) and
(\ref{eq16}) be
\begin{equation}
\left(\frac{A'{}^2}{AB}\right)'-\frac{4}{\ell^2}
A'=0\,.
\label{eq18}
\end{equation}
We find that Eq.~(\ref{eq18}) are derived from Eq.~(\ref{eq17}).
This fact will yield a new result with $\Lambda \neq 0$.
The solutions with and without a cosmological constant will be exhibited
in the next section.

%%%%%%%%%%%%%%%%%%%%%%%%%%%%%%%%%%%%%%%%%%%%%%%%%%%%%%%%%%%%%%%%%%
\section{The solution of the field equation
\label{sec3}}
%%%%%%%%%%%%%%%%%%%%%%%%%%%%%%%%%%%%%%%%%%%%%%%%%%%%%%%%%%%%%%%%%%

%%%%%%%%%%%%%%%%%%%%%%%%%%%%
\subsection{$\Lambda=0 (\ell \rightarrow \infty)$ case}
%%%%%%%%%%%%%%%%%%%%%%%%%%%%
In this subsection, we focus on the $\Lambda=0$ case.
Then, it is shown that the same result as Ref.~\cite{MH} is re-derived.
If $\Lambda=0$, Eq.~(\ref{eq17}) turns out to be
\begin{equation}
B=\frac{A'{}^2}{2\sqrt{2}\alpha E_0}\frac{1}{A}\,.
\label{eq19}
\end{equation}
Let us take the following new coordinate:
\begin{equation}
\bar{r}=\int\sqrt{AB} dr\,.
\end{equation}
Using the new coordinate, the metric (\ref{eq2}) takes the form
\begin{equation}
ds^2=-f(\bar{r})dt^2+\frac{d\bar{r}^2}{f(\bar{r})}+f(\bar{r})d\theta^2\,,
\end{equation}
where $f(\bar{r})\equiv A(r(\bar{r}))$.
This implies that it can be taken as $AB=1$ without a loss of generality.
When $AB=1$, Eq.~(\ref{eq19}) turns out to be
\begin{equation}
A'=\sqrt{2\sqrt{2}\alpha E_0}\,.
\end{equation}
This equation can easily be solved and the solution is
\begin{equation}
A=\xi r+C\,,
\end{equation}
where $C$ is an integration constant and $\xi\equiv\sqrt{2\sqrt{2}\alpha
E_0}$. Then, the metric becomes
\begin{equation}
ds^2=-(\xi r+C) dt^2+\frac{dr^2}{\xi r+C}+(\xi r+C)d\theta^2\,.
\end{equation}
Since the constant $C$ can be removed by the transformation
$r\rightarrow r-\frac{C}{\xi}$, we obtain
\begin{equation}
ds^2=-\xi r dt^2+\frac{dr^2}{\xi r}+\xi r d\theta^2\,.
\label{eq25}
\end{equation}
In order to compare this with the result of Ref.~\cite{SS}, we introduce
new variables:
\begin{equation}
r=\frac{\xi}{4}\bar{r}^2\,,\quad
\theta=\frac{2\bar{\theta}}{\xi}\,,\quad
t=\frac{2\bar{t}}{\xi}\,.
\end{equation}
Consequently, the solution in the $2+1$ dimensional gravity coupled to
NED is given by
\begin{equation}
ds^2=-\bar{r}^2 dt^2+d\bar{r}^2+\bar{r}^2 d\bar{\theta}^2\,.
\label{eq27}
\end{equation}
This form of the metric is the same as in Ref.~\cite{SS}.
Moreover, the solution (\ref{eq27}) is not a ``usual'' black hole,
but a singularity arises at $\bar{r}=0$; so it should be called ``black
point'' (BP). For example, in Ref.~\cite{Sol}, BP appears in $3+1$
dimensions when a logarithmic $U(1)$ gauge theory couples to gravity. In
Ref.~\cite{HW}, the case with not only $U(1)$ theory but also a
dilatonic field is studied and a similar result has been found.
Of course, this solution agrees with that in Ref.~\cite{SS}.

%%%%%%%%%%%%%%%%%%%%%%%%%%%%
\subsection{$\Lambda\ne 0$ case}
%%%%%%%%%%%%%%%%%%%%%%%%%%%%
In this subsection, we consider the field equation (\ref{eq17}) for a
$\Lambda\ne 0$ case. Again, without a loss of generality, it can be taken
as $AB=1$ (see subsection 3.1). Then, Eq.~(\ref{eq17}) becomes
\begin{equation}
(A')^2-\frac{4}{\ell^2}
A=\frac{4\alpha E_0}{\sqrt{2}}\,.
\label{eq28}
\end{equation}
This differential equation can be solved analytically, and the general
solution is
\begin{equation}
\xi^2+\frac{4}{\ell^2} A=\frac{4}{\ell^4}(r-C)^2\,,
\label{eq29}
\end{equation}
where $C$ is an integration constant and we
define $\xi\equiv\sqrt{2\sqrt{2}\alpha E_0}$, again.
When $\Lambda$ approaches to zero, the solution (\ref{eq29}) should
coincide with (\ref{eq25}). This condition implies that
$C=\xi/(2\Lambda)$. Then, the metric reduces to
\begin{equation}
ds^2=-(\xi r+\frac{r^2}{\ell^2})dt^2+\frac{dr^2}{\xi r+r^2/\ell^2}
+(\xi r+\frac{r^2}{\ell^2})d\theta^2\,.
\label{eq30}
\end{equation}
Let us take the following variables
\begin{equation}
r+\frac{1}{\xi}\frac{r^2}{\ell^2}=\frac{\xi}{4}\bar{r}^2\,,\quad
\theta=\frac{2\bar{\theta}}{\xi}\,,\quad
t=\frac{2\bar{t}}{\xi}\,.
\end{equation}
Using them, the metric (\ref{eq30}) can be rewritten as
\begin{equation}
ds^2=-\bar{r}^2d\bar{t}^2+\frac{d\bar{r}^2}{1+\bar{r}^2/\ell^2}
+\bar{r}^2 d\bar{\theta}^2\,.
\label{eq32}
\end{equation}
This is the new solution of $2+1$ gravity coupled with NED. Of course, it
can be seen when $\Lambda$ reaches zero, the metric (\ref{eq32}) becomes
(\ref{eq27}). Moreover, in the solution (\ref{eq32}), we notice that if
the spatial part of the metric reduce $1+(r^2/\ell^2) \rightarrow
(r^2/\ell^2)-M$,it seems to be a ``black hole'' solution. In fact, we
introduce new variables as
\begin{equation}
\bar{r}=L\tilde{r}\,,\quad\bar{\theta}=\frac{\tilde{\theta}}{L}\,,\quad
t=\frac{\tilde{t}}{L}\,,
\label{mass}
\end{equation}
where $L$ is some arbitrarily constant. Then, the metric (\ref{eq32}) reduce to
\begin{equation}
ds^2=-\tilde{r}^2d\tilde{t}^2+\frac{d\tilde{r}^2}{1/L^2+\tilde{r}^2/\ell^2}
+\tilde{r}^2d\tilde{\theta}^2
\label{met}
\end{equation}
Finally, let us take a constant $L$ as $M=-1/L^2$, we get 
\begin{equation}
ds^2=-r^2dt^2+\frac{dr^2}{(r^2/\ell^2)-M}+r^2d\theta^2\,.
\label{eq33}
\end{equation}
Where $\tilde{t}$, $\tilde{r}$, $\tilde{\theta}$ are replaced by $t$, $r$,
$\theta$ for simplicity.
The metric has the same form of the ansatz (\ref{eq2}), so  the solution (\ref{eq33})
must satisfy the field equation (\ref{eq14}),(\ref{eq15}) and (\ref{eq16}).
It can be shown the solution (\ref{eq33}) satisfy (\ref{eq14}) and (\ref{eq16})
easily. Furthermore, (\ref{eq15}) becomes
\begin{equation}
\frac{4}{4r^2} \left( \frac{r^2}{\ell^2}-M \right) -\frac{1}{\ell^2}
=\frac{\alpha}{2}\sqrt{\frac{2E_0^2}{r^4}}\,.
\label{eq34}
\end{equation}
This equation means that coupling constant $\alpha$ or $E_0$ must be negative
if the ``black hole mass'' $M$ is lager than zero, and the ``mass'' is
\begin{equation}
M=-\frac{\alpha E_0}{\sqrt{2}}\,.
\label{eq34a}
\end{equation}
So we conclude the meaning of ``black hole mass'' of the
space time (\ref{eq33}) is different from the BTZ black hole mass,
because $M$ is determined  by the coupling constant $\alpha$ and $E_0$.

``Black hole'' solution (\ref{eq33}) has the horizon at
\begin{equation}
r_H=\sqrt{M}\ell\,.
\label{eq34b}
\end{equation}
This point is not a singular point, because the Kretchmann invariant
${\cal R} = R_{\mu\nu\lambda\sigma}R^{\mu\nu\lambda\sigma}$ 
from the metric (\ref{eq33})
\begin{equation}
{\cal R} = \frac{8}{\ell^4} + \frac{4}{r^4}
\left( M - \frac{r^2}{\ell^2} \right) ^2
\,,
\label{eq334}
\end{equation}
does not diverge at $r_H$. However, ${\cal R}$ diverge 
at $r=0$, this implies that space-time (\ref{eq33}) has a
singular point at $r=0$. 

%%%%%%%%%%%%%%%%%%%%%%%%%%%%%%%%%%%%%%%%%%%%%%%%%%%%%%%%%%%%%%%%%%
\section{Radial null rays
\label{sec4}}
%%%%%%%%%%%%%%%%%%%%%%%%%%%%%%%%%%%%%%%%%%%%%%%%%%%%%%%%%%%%%%%%%%
In the previous section, we get the new solution (\ref{eq33}) in $2+1$
gravity. Then we wonder that, a solution (\ref{eq33}) describe
black hole or not? That is why we consider radial null rays in this
section to answer the above question.

\begin{figure}[htbp]
\begin{center}
\includegraphics[width=100mm]{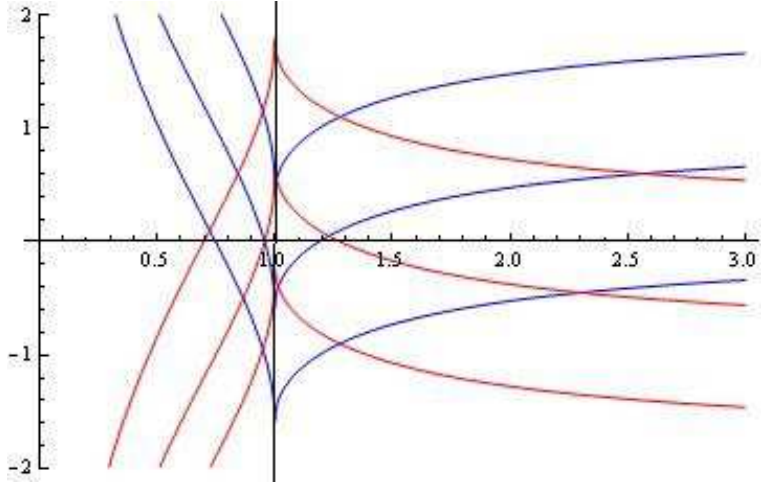}
\end{center}
%%%%%%%%%%%%%%%%%%%%%%%%%%%%%%%%%%%%%%%%%%%%%%%%%%%%%%%%%%%%%%%%%%
\caption{Radial null rays in space-time (\ref{eq33}). The horizontal
axis represents the particle position $r$, and the vertical axis
represents the coordinate time $t$.We set $\ell=1$,$\sqrt{M}=1$ for
simplicity. The black solid line represents the horizon $r=r_H$.
Outgoing rays are described by the blue lines, and incoming rays are
red lines.}
%%%%%%%%%%%%%%%%%%%%%%%%%%%%%%%%%%%%%%%%%%%%%%%%%%%%%%%%%%%%%%%%%%
\label{fig1}
\end{figure}

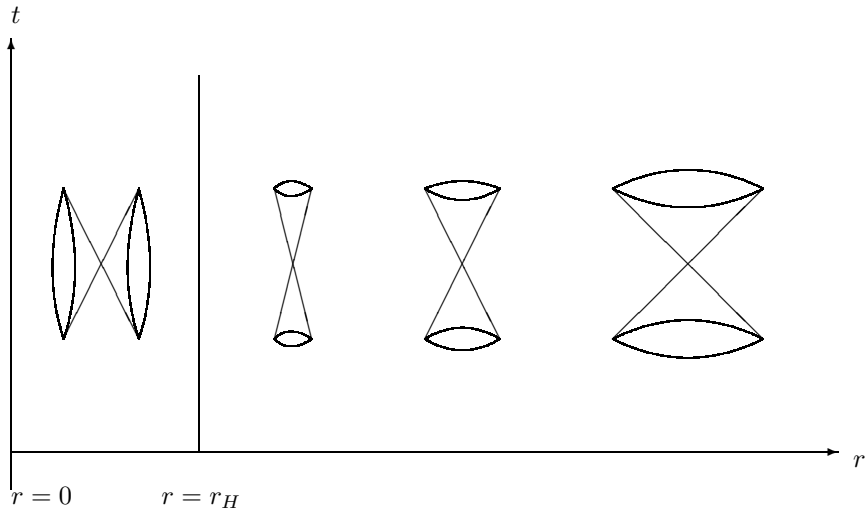
\begin{figure}[htbp]
\unitlength=1mm
\begin{picture}(120,70)

\put(0,10){\vector(1,0){110}}
\put(0,5){\vector(0,1){60}}
\put(25,10){\line(0,1){50}}

\put(80,45){\line(1,-1){20}}
\put(80,25){\line(1,1){20}}
\qbezier(80,45)(90,50)(100,45)
\qbezier(80,45)(90,40)(100,45)
\qbezier(80,25)(90,20)(100,25)
\qbezier(80,25)(90,30)(100,25)

\put(55,45){\line(1,-2){10}}
\put(55,25){\line(1,2){10}}
\qbezier(55,45)(60,47)(65,45)
\qbezier(55,45)(60,42)(65,45)
\qbezier(55,25)(60,28)(65,25)
\qbezier(55,25)(60,22)(65,25)

\put(35,45){\line(1,-4){5}}
\put(35,25){\line(1,4){5}}
\qbezier(35,45)(37,47)(40,45)
\qbezier(35,45)(37,43)(40,45)
\qbezier(35,25)(37,27)(40,25)
\qbezier(35,25)(37,23)(40,25)

\put(7,45){\line(1,-2){10}}
\put(7,25){\line(1,2){10}}
\qbezier(7,45)(4,34)(7,25)
\qbezier(7,45)(10,34)(7,25)
\qbezier(17,45)(20,34)(17,25)
\qbezier(17,45)(14,34)(17,25)

\put(20,3){$r=r_H$}
\put(0,3){$r=0$}
\put(112,8){$r$}
\put(0,67){$t$}

\end{picture}
\caption{%
 The behaviour of light cones of space-time (\ref{eq33}) in the 
polar coordinates ($r,t$).}
%%%%%%%%%%%%%%%%%%%%%%%%%%%%%%%%%%%%%%%%%%%%%%%%%%%%%%%%%%%%%%%%%%
\label{fig2}
\end{figure}

A radial world line for a photon in space-time (\ref{eq33}) is
\begin{equation}
ds^2=-r^2dt^2+\frac{dr^2}{(r^2/\ell^2)-M}=0\,.
\label{eq35}
\end{equation}
We then have
\begin{equation}
dt = \pm \frac{\ell}{r}\frac{dr}{\sqrt{r^2-r^2_H}}\,.
\label{eq36}
\end{equation}
Where $+$ sign stands for the outgoing, and the $-$ sign
for the incoming.
This equation can integrate analytically, and the solution is
\begin{equation}
t+t_0
=\begin{cases}\pm\frac{1}{\sqrt{M}}\arctan\left(
\frac{\sqrt{r^2/r^2_H-1}}{1-r^2/r^2_H} \right) & \text{($r>r_H$)} \\
\pm\frac{1}{\sqrt{M}}\ln\left( \frac{1+\sqrt{1-r^2/r_H^2}}{1-\sqrt
{1-r^2/r_H^2}} \right) & \text{($r<r_H$)} \,.
\end{cases}
\label{eq37}
\end{equation}
Light like geodesic described by Eqs.~(\ref{eq37}) in polar
coordinates is shown in Fig.\ref{fig1}. From Fig.\ref{fig1},
light cones in the space-time (\ref{eq33}) behave like Fig.\ref{fig2}.
This implies that light cones spread out when $r$ reaches to
infinity. Action (\ref{eq1}) represent asymptotically AdS
space-time, this result is consistent. On the other hand, when $r$
reaches to horizon $r_H$, light cones become narrower. At the
horizon $r=r_H$, the slope $dt/dr$ reaches to infinity, so light
cones collapse. Next, we attend light cones that lies inside
the horizon. Light cones are tipped over and null rays go to the
singular point $r=0$. This implies that any particle (except tachyon)
can not escape from that lies in side of the horizon.

Above behavior of space-time (\ref{eq33}) is similar to the well
known black holes (For example, Schwarzschild black hole in AdS
space-time). That is why we suggest our new solution (\ref{eq33})
describes one of the black hole space-time.

%%%%%%%%%%%%%%%%%%%%%%%%%%Figure1%%%%%%%%%%%%%%%%%%%%%%%%%%%%%%%%%
%*****************************************************************
%*****************************************************************

%%%%%%%%%%%%%%%%%%%%%%%%%%%%%%%%%%%%%%%%%%%%%%%%%%%%%%%%%%%%%%%%%%
%%%Figure2
%%%%%%%%%%%%%%%%%%%%%%%%%%%%%%%%%%%%%%%%%%%%%%%%%%%%%%%%%%%%%%%%%%
%%%%%%%%%%%%%%%%%%%%%%%%%%%%%%%%%%%%%%%%%%%%%%%%%%%%%%%%%%%%%%%%%%

\newpage

%%%%%%%%%%%%%%%%%%%%%%%%%%%%%%%%%%%%%%%%%%%%%%%%%%%%%%%%%%%%%%%%%%
\section{Conclusion and remarks
\label{sec5}}
%%%%%%%%%%%%%%%%%%%%%%%%%%%%%%%%%%%%%%%%%%%%%%%%%%%%%%%%%%%%%%%%%%
In this paper, the solution of the field equation for $2+1$ dimensional
gravity coupled to NED has been drawn. In the $\Lambda=0$ case, it agrees
with the solution derived in Ref.~\cite{MH}. On the hand, in the
$\Lambda\ne 0$ case, we have obtained the new solution (\ref{eq32}).
We also get the new black hole solution (\ref{eq33}) from the field
equation (\ref{eq14}),({\ref{eq15}) and (\ref{eq16}). We notice that
the black hole mass $M$ is related to the coupling constant
$\alpha$ and the amplitude of the field strength $E_0$.
Addition, the new solution (\ref{eq33}) has the horizon at $r=\sqrt{M}
\ell$. However, from the Kretchamann invariant, physical singular
point of this space-time lies at $r=0$, only.
We also investigate radial null rays of the space-time (\ref{eq33}),
and the feature of rays are similar to Schwarzschild one.
That is why we suggest that the new solution (\ref{eq33}) from the
action (\ref{eq1}) describe one of the black hole solution.

By the way, although the sign in front of $dr^2$ is reversed by
$r<r_c$ in the line element (\ref{eq33}), the sign
of the coefficient of $dt^2$ is unchanged. 
Thus it seems that the $r=r_c$ is not a genuine horizon.
The singular behavior should be further investigated with an interest.

Moreover,  in the BTZ black hole, theormodynamical consideration 
has been performed and the behavior of thermodynamic quantities are
investigated.
The space-time described the solution (\ref{eq33}) should be studied
along with the similar thermodynamical context. It is a future issue to
survey thermodynamics of the space-time.

%%%%%%%%%%%%%%%%%%%%%%%%%%%%%%%%%%%%%%%%%%%%%%%%%%%%%%%%%%%%%%%%%%
\section*{Acknowledgments}
%%%%%%%%%%%%%%%%%%%%%%%%%%%%%%%%%%%%%%%%%%%%%%%%%%%%%%%%%%%%%%%%%%

%\acknowledgments
%%%%%%%%%%%%%%%%%%%%%%%%%%%%%%%%%%%%%%%%%%%%%%%%%%%%%%%%%%%%%%%%%%%%%%%%%%%
%Acknowledgements
%%%%%%%%%%%%%%%%%%%%%%%%%%%%%%%%%%%%%%%%%%%%%%%%%%%%%%%%%%%%%%%%%%%%%%%%%%%
%\begin{acknowledgments}
I would like to thank Doctor K.~Shiraishi for helpful discussions.
I also thank him for reading the manuscript.
%\end{acknowledgments}
%%%%%%%%%%%%%%%%%%%%%%%%%%%%%%%%%%%%%%%%%%%%%%%%%%%%%%%%%%%%%%%%%%%%%%

%%%%%%%%%%%%%%%%%%%%%%%%%%%%%%%%%%%%%%%%%%%%%%%%%%%%%%%%%%%%%%%%%%%%%%%%%%%
%thebibliography
%%%%%%%%%%%%%%%%%%%%%%%%%%%%%%%%%%%%%%%%%%%%%%%%%%%%%%%%%%%%%%%%%%%%%%%%%%%

%\bibliographystyle{apsrev}
\bibliographystyle{apsrev4-1}
%\bibliography{}

%%%%%%%%%%%%%%%%%%%%%%%%%%%%%%%%%%%%%%%%%%%%%%%%%%%%%%%%%%%%%%%%%%%%%%
%%%References
%%%%%%%%%%%%%%%%%%%%%%%%%%%%%%%%%%%%%%%%%%%%%%%%%%%%%%%%%%%%%%%%%%%%%%

%%%%%%%%%%%%%%%%%%%%%%%%%%%%%%%%%%%%%%%%%%%%%
\end{document}